\documentclass[cits]{JINST}

\usepackage{units}
\usepackage{upgreek}
\usepackage{url}

\renewcommand{\mu}{\upmu}
\renewcommand{\nu}{\upnu}
\renewcommand{\theta}{\uptheta}
\renewcommand{\phi}{\upphi}
\renewcommand{\pi}{\uppi}
\renewcommand{\xi}{\upxi}
\renewcommand{\delta}{\updelta}
\renewcommand{\gamma}{\upgamma}
\renewcommand{\beta}{\upbeta}
\renewcommand{\kappa}{\upkappa}
\renewcommand{\tau}{\uptau}
\renewcommand{\omega}{\upomega}
\renewcommand{\rho}{\uprho}
\renewcommand{\alpha}{\upalpha}
\renewcommand{\epsilon}{\upepsilon}

\newcommand{\mr}[1]{{\mathrm{#1}}}
\graphicspath{{figures/}}

\title{A Novel Self-supporting GEM-based Amplification Structure for a Time Projection Chamber at the ILC}

\author{Ties Behnke$^a$, Ralf Diener$^a$, Christoph Rosemann$^a$, Lea Steder$^a$\thanks{Corresponding
Author}\\
\llap{$^a$}DESY,\\
  Notkestrasse 85, 22607 Hamburg, Germany\\
  E-mail: \email{lea.steder@desy.de}}

\abstract{Modern time projection chambers are increasingly based on micro pattern gas detector readout systems.
In this paper  a self-supporting method used to mount Gas Electron Multiplier foils is presented. It is based on light weight ceramic grids, and promises to cover large readout areas with minimum dead zones and material, while ensuring a flat and mechanically stable mounting. 
The structure has been tested in a Time Projection Chamber prototype, using cosmic muon tracks.
The impact of the mounting structure on the charge measurement, the track reconstruction and the single point resolution is quantified.}


\begin{document}

\section{Introduction}
Time Projection Chambers (TPCs) are discussed as tracking detectors for a number of projects in nuclear and particle physics.
They are proposed because TPCs combine good spatial resolution with excellent, robust and efficient three-dimensional pattern recognition and the possibility to identify particles via their specific energy loss, $dE/dx$.
Modern TPC systems are increasingly designed based on micro-pattern gas detectors (MPGD) for readout.
These systems offer a number of potential advantages compared to the traditional wire-chamber based readout.
MPGDs have amplification structures which are of a size ($\mathcal{O}$(100)~$\mu$m) similar to the resolution anticipated.
Effects connected to the length scale in the amplification ---as, for example, distortions introduced by drifting electrons transverse to the magnetic field--- scale as the feature size and are therefore much reduced in MPGD based systems.
Another advantage of MPGDs is that they suppress the backdrift of ions into the drift volume. This feature allows the construction of TPCs which can operate continuously. This is however ---while essential for the application of MPGD in many instances--- not further discussed in this paper.
Here, a GEM (Gas Electron Multiplier) \cite{gem} based TPC system is described.

A typical GEM based readout system for a TPC consists of two or more GEM foils stacked on top of a readout pad plane.
Electrons produced in the volume of the TPC are drifted towards the readout.
Gas amplification takes place at each GEM.
Per GEM an amplification of typically a factor of 100-200 is possible resulting in a total system amplification between $10^3$ and $10^4$.
The performance of the system depends on the intrinsic amplification properties of the GEM, and on the mechanical properties like distances between GEMs, flatness of the GEM foils and stability of the overall system.

A technical problem for GEM systems is the mechanical support of a large scale amplification system on the end plate.
The amplification system should cover the end plate without dead areas, should be mechanically stable and robust, and should minimize the amount of dead material introduced.
The GEM foil should be supported stably, kept at a constant distance from the readout pad plane or its neighbor GEM, and not move under the influence of the electric or magnetic field.
A number of different solutions to this problem have been published over the years \cite{heraB, panda, fair} including some which use also thin spacers between GEMs.
However the solution proposed here is special insofar as there are no thicker spacers on the edges of the GEM system, as it is for example the case with the COMPASS GEM support system \cite{compass}.

In this paper, a system is proposed to support the GEM foils by a ceramic grid. 
The system was originally developed for the TPC for the ILD (International Large Detector) experiment proposal at the ILC, but can be applied to other systems as well. 
The grid serves a dual function.
When glued to the GEM a very stable, light-weight mechanical system is formed, with a very high bending strength, similar to the way modern sandwich materials are based on sheets of honeycomb material with thin skins on the top and the bottom.
By properly adjusting the mesh size of the grid  to the typical length scale by which GEM foils undulate, the GEM foils can be flattened without applying undue mechanical stress.

The design parameters and properties of the proposed system are evaluated based on GEM foils of $10 \times 10$~cm$^2$.
In the following, the design of the system is described, and its performance is evaluated.

\section{Mechanical Design of the Grid GEM}
\label{sec:prod}
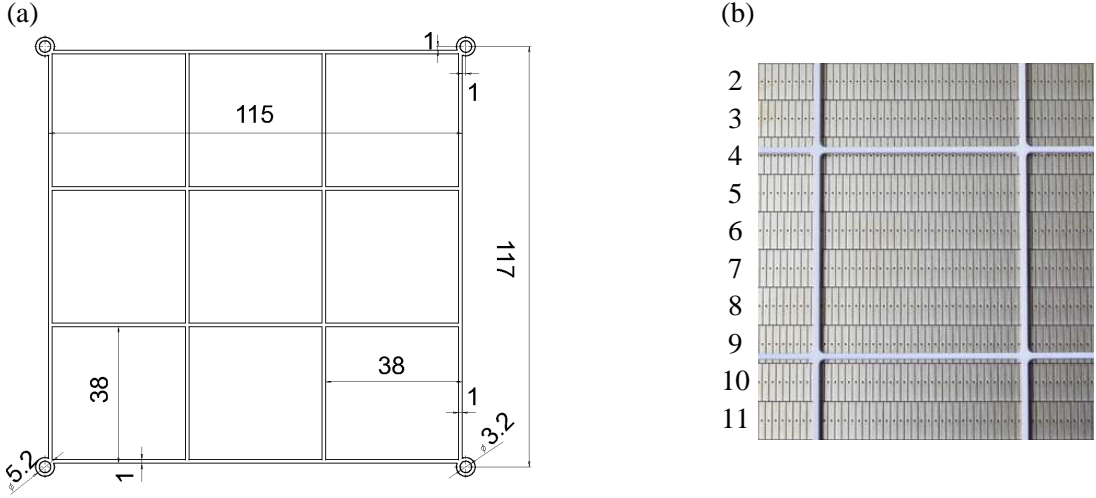
\begin{figure}
\unitlength1cm
\centering
\begin{picture}(15,6.5)
\put(0,0){\includegraphics[width=7cm,clip=]{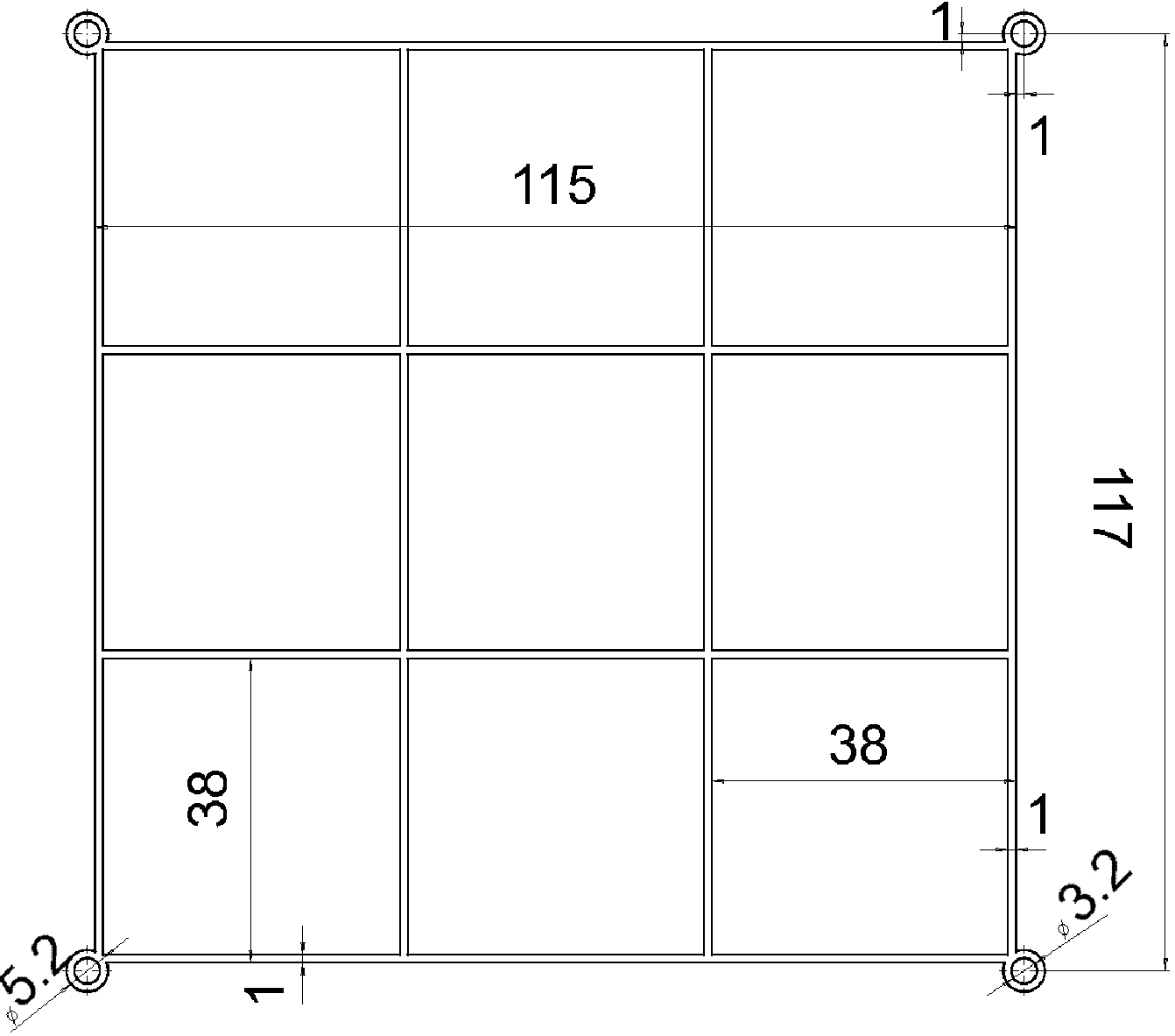}}
\put(10,0.75){\includegraphics[height=5cm,clip=]{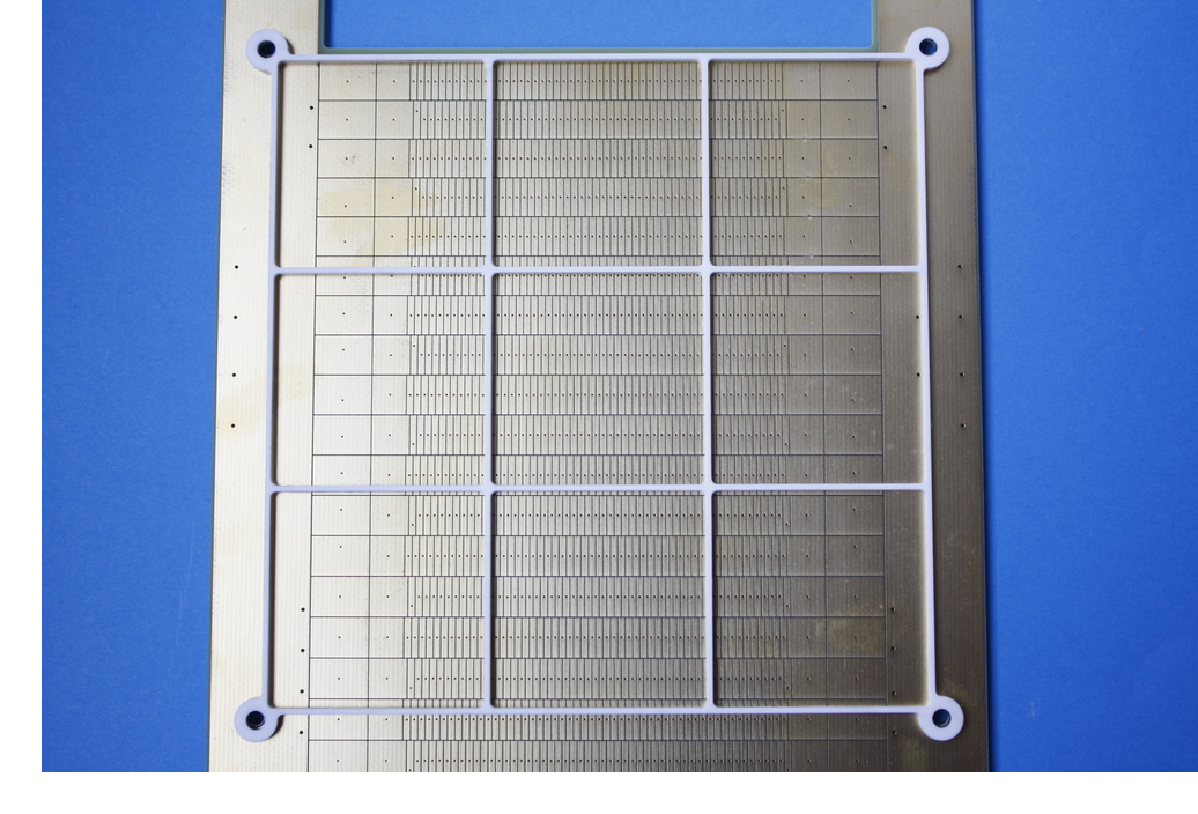}}
\put(0,6.3){(a)}
\put(9.5,6.3){(b)}
\put(9.5,0.9){11}
\put(9.5,1.4){10}
\put(9.5,1.9){ 9}
\put(9.5,2.4){ 8}
\put(9.5,2.9){ 7}
\put(9.5,3.4){ 6}
\put(9.5,3.9){ 5}
\put(9.5,4.4){ 4}
\put(9.5,4.9){ 3}
\put(9.5,5.4){ 2}
\end{picture}
\caption{(a) Drawing of ceramic grid.
All measures are given in millimeters.
(b) Relative position of grid and pad plane.
Row one and twelve are also in the sensitive area of the GEM foils, but not connected due to the limited number of channels or the readout electronics.}
\label{fig:grid}
\end{figure}

The central part of the proposed GEM system is a stack of GEM foils separated and supported by ceramic grids. 
The grid is made of an aluminum oxide ceramic ($\mr{Al_2 O_3}$) \cite{ANC}.
This material is very stiff, an excellent insulator and machinable by laser cutting.
The support is designed as a grid with a mesh size of about \unit[38]{mm}.
The bars of the grid are \unit[1]{mm} wide, which is the minimum width which currently is possible in this technology (see figure~\ref{fig:grid} for a technical drawing and list of specifications). 
A basic unit consists of a grid glued on both sides to a GEM foil producing a very stable and light-weight sandwich.
Here, a system of 'GEM - two grids - GEM - two grids - GEM' is used for a triple grid GEM stack, which is mounted on the readout plane. 
Such a stack provides \unit[2]{mm} wide transfer gaps.
Several basic units can be combined in different ways and with additional or less grids as spacers, so that a very flexible system exists to build up a stack of GEMs on top of a readout plane, with different distances between the different units.
The basic unit is produced by pulling the GEM foils on a flat vacuum table and then, using a proper mounting rig, positioning and glueing the GEM to the grid.
A robot arm is used to meter the appropriate amounts of glue on the grid bars (a two component epoxy, Polybond EP 4619/3~\cite{PT}).
Currently, only the outer bars of the grid are glued to the GEM.
This is to avoid that glue leaks into the GEM holes in areas where the grid bars cover the active area of the GEM.
In the future GEM foils will be specially produced which will provide narrow gluing surfaces without holes underneath the grid bars.
It should be noted that the full mechanical strength will only be reached with such GEM foils.  
%%%%%%%%%%%%%%%%%%%%%%%%%%%%%%%%%%%%%%%%%%%%%%%%%%%%%%%%%%%%%%
%%%%%%%%%%%%%%%%%%%%%%%%%%%%%%%%%%%%%%%%%%%%%%%%%%%%%%%%%%%%%%
A technical drawing of the grid is shown in figure \ref{fig:grid}(a).
The relevant characteristics are summarized in table \ref{tab:grid}.
%%%%%%%%%%%%%%%%%%%%%%%%%%%%%%%%%%%%%%%%%%%%%%%%%%%%%%%%%%%%%%
\begin{table}[h]
\caption{Material properties and dimensions of grid support structure.}
\centering
\frame{
\begin{tabular}{l l}
\multicolumn{2}{c}{grid support measures}\\
\hline
material & $\mr{Al_2 O_3}$\\
radiation length & $\mr{X_0 =\unit[7.0]{cm}}$\\
resistivity & $\mr{> \unit[10^{12}]{\Omega cm}}$\\
bending strength & $\mr{\sigma_B = \unit[350]{MPa}}$\\
outer dimensions & $\mr{\unit[117 \times 117]{mm^2}}$\\
cell size & $\mr{\unit[37 \times 37]{mm^2}}$\\
sensitive GEM area & $\mr{\unit[10 \times 10]{cm^2}}$\\
optical transparency & \unit[96]{\%}\\
structure width & \unit[1]{mm}\\
structure height & \unit[1]{mm}\\
\end{tabular}
}
\label{tab:grid}
\end{table}
%%%%%%%%%%%%%%%%%%%%%%%%%%%%%%%%%%%%%%%%%%%%%%%%%%%%%%%%%%%%%%

For the prototype system the grid covers about \unit[4]{\%} of the active area of the GEM foil.
This is significantly less than in more conventional systems, where typically a stable frame external to the GEM takes the stress from stretching the GEM over the readout plane.
Using naively the standard GEM provided mounting frames made from \unit[1]{cm} wide strips of glass fiber reinforced plastic and scaling it up, the support would take about \unit[35]{\%} of the total area.

The flatness of the GEM foils is another key parameter for their performance.
The gain of a GEM depends on the potential applied across the GEM, and on the field on the top and at the bottom of the GEM.
Changing the distance between the GEM and, e.g., the readout pad plane, will change locally the electric field and thus the gain.
Therefore, for optimal performance a flat mounting is important.
In addition, changes in the distance will create potentially areas with high fields which might be more sensitive to breakdown or instabilities.
Studies done in simulation and validated with experimental measurements \cite{lea} have shown that fluctuations of the transfer distances between the GEMs ---due to height deviations of the GEM foils of the order of \unit[600]{$\mu$m}--- result in variations of the energy resolution of up to \unit[5]{\%}, which would still allow for a significant contribution to the particle identification via $dE/dx$ measurements in the TPC.
The spatial resolution is to first order not affected by this height variations of the foils (see section \ref{sec:track}).

To evaluate the feasibility and performance of the system, a triple GEM amplification structure was built.
It is placed on a readout plane instrumented with pads with a pitch of \unit[1.27~$\times$~7]{mm$^2$}, shown in figure \ref{fig:grid}(b).
Adjacent rows of pads are staggered by half a pad pitch and the grid is aligned to the pads.
For this photo, the grid has been placed directly onto the readout pad plane. About \unit[26]\% of the pads are shadowed by the grid.
A key objective therefore is to study the impact this shadowing has on the performance, overall and especially for those pads directly affected by the grid. 

The assembled system was mechanically measured and the maximum deviation from a flat plane was found to be about \unit[500]{$\mu$m}. 
Its impact on the point resolution is sufficiently small to be neglected and expected to meet the envisaged particle identification capabilities via $dE/dx$ measurements.

The described system has been tested inside a prototype TPC, with and without magnetic field.
It allows for stable operation using the same gas flow rate and high voltage parameters as for GEM stacks operated without a grid structure.
No operational problems like charge-up or instabilities of the GEM foils were found over extended running periods of several weeks.

\section{Experimental Setup to Study a Grid GEM TPC with Cosmic Muon Data}
\label{sec:setup}
The triple grid GEM stack composed of \unit[10~$\times$~10]{cm$^2$} GEMs (double conical holes, hole pitch $\mr{\unit[140]{\mu m}}$, \unit[50/70]{$\mu$m} inner/outer hole diameter) was operated in a prototype TPC with an active area of $\unit[27]{cm^2}$ and a drift length of \unit[66]{cm} at magnetic fields of up to \unit[4]{T}.
The GEM support structure was mounted on top of the readout plane described in section \ref{sec:prod}.
Ten complete pad rows were read out with charge sensitive preamplifiers followed by a \unit[12.5]{MHz} flash ADC system \cite{daq}.
As counting gas, a mixture of \unit[95]{\%} Argon and \unit[5]{\%} Methane was used.
The drift field was set to \unit[90]{V/cm} in order to be ---due to the velocity plateau for this gas at this value--- fairly independent from small drift field variations.
The GEMs were used with voltages across them ranging between \unit[320]{V} and \unit[325]{V}.
Fields of \unit[1.5]{kV/cm} were applied in the \unit[2]{mm} wide transfer gaps, while a field of \unit[3]{kV/cm} was used in the \unit[3]{mm} wide induction region.
With the current TPC prototype setup, an absolute gain measurement is not possible. Using the parametrization of a GEM based system described in \cite{sven}, a gain of 10,000 has been estimated for the setup.
Two scintillator counters ---above and below the prototype--- operated in coincidence were used to trigger on cosmic muons. 
%More details about the trigger system and the whole setup can be found in \cite{thorsten}.

The coordinate system used in the reconstruction is defined by the pad plane ---x is pointing horizontally over the pad rows, while y follows the vertical columns--- and the drift distance along the chamber axis, corresponding to the z axis.

The reconstruction of the recorded data is divided into three steps.
First, the pad-wise charge deposition from different time bins (corresponding to z values) is recorded. Only hits which cross at the rising edge a threshold set at about twice the pedestal value are used. The hit reconstruction in time is terminated if the charge crosses a second, lower threshold. Since one hit is expected to extend across more than one pad, pads are then clustered row-wise into hits.
A simple center of gravity method is used to determine the coordinates of this hit across the rows. The time of the reconstructed hit is calculated by looking at the distribution of the differences between bins which follow in time. The maximum of this distribution corresponding to the inflection point of the pulse in the rising edge, is used as an estimator for the time. %-spatial, and charge.
Second, a track finding algorithm is applied, which assigns the hits to a track candidate.
Finally, the tracks are fitted with an algorithm assuming a circular path.
More details about the reconstruction algorithms can be found in~\cite{matthias}.

\section{Grid Impact on Charge Measurement}
\begin{figure}
\unitlength1cm
\centering
\begin{picture}(15,6)
 \put(0.1,0.22){\includegraphics[width=8.5cm,clip=]{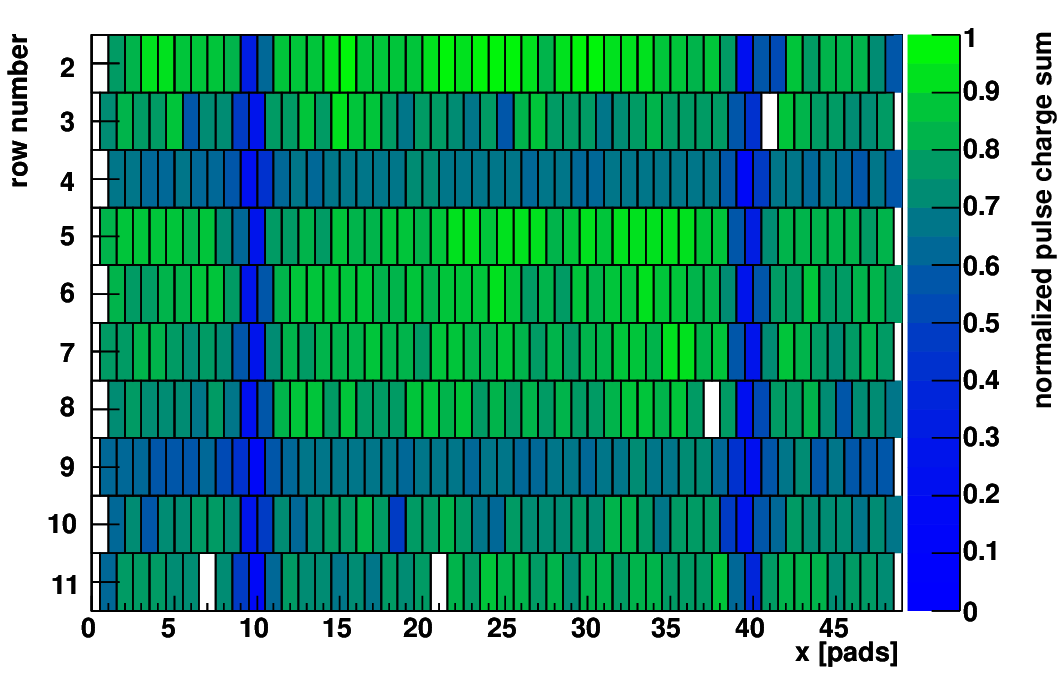}}
 \put(9,0.15){\includegraphics[width=5.8cm,clip=]{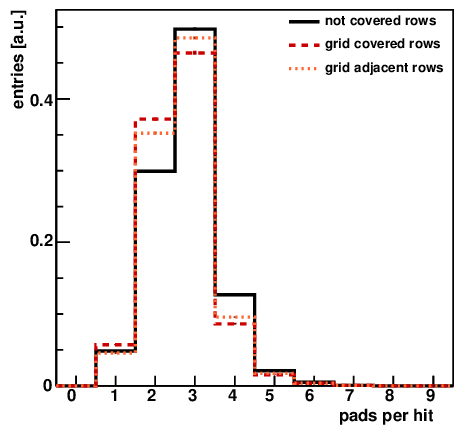}}
 \put(0.1,5.7){(a)}
 \put(9,5.7){(b)}
\end{picture}
\caption{(a) Total charge ---integrated over a measurement run, and normalized to the total number of triggers in each bin --- on the readout pads.
%In each row, pads 9, 10 and pads 39, 40 are covered partly by the vertical grid bars.
%The horizontal bars are positioned over row four and nine of the pad plane.
(b) Influence of the horizontal grid bar on the number of pads contributing to a hit.
%The distribution has been normalized to the total number of entries.
In both figures, data recorded at a magnetic field of \unit[4]{T} are shown.}
\label{fig:Qsumhhit}
\end{figure}

Since parts of the grid shadow some pads, an impact on the reconstructed hit parameters for hits on these pads is expected. 
The impact of the grid on the charge has been studied by investigating the total amount of charge deposited throughout a long data taking period with about 61,000 triggers on different pads.
The cosmic rays illuminate the sensitive area uniformly, such that deviations from an uniform charge distribution are likely to be caused by the grid structure.\\ %Here, E$\times$B effects are negligible, since the sensitive area is in all directions more than \unit[2]{cm} away from the ground potential at the gap around the GEM structure.\\
In figure \ref{fig:Qsumhhit}(a) the total recorded charge per pad integrated over the measurement run is shown. Each bin has been normalized to the total number of triggers recorded in this bin. 
%%%%%%%%%%%%%%%%%%%%%%%%%%%%%%%%%%%%%%%%%%%%%%%%%%%%%%%%%%%%%%
%%%%%%%%%%%%%%%%%%%%%%%%%%%%%%%%%%%%%%%%%%%%%%%%%%%%%%%%%%%%%%
The structure of the grid is clearly visible through regions of reduced overall charge.
Since the pads are highly elongated, 
%being only \unit[1.27]{mm} wide, but \unit[7]{mm} high,
the impact is expected to be different for vertical and horizontal bars. 
The vertical bars, going in the direction of increasing row numbers, cover on average close to \unit[50]{\%} of a pad in this region.
For the horizontal bars, oriented parallel to the x axis, the bars cover only  about \unit[15]{\%}.

The observed reduction in charge per pad is about \unit[60]{\%} for the vertical bars and \unit[25]{\%} for the horizontal bars, which is in rough agreement with the assumption that the charge reduction is to the first order proportional to the geometrical area shadowed by the grid.\\
The results show a clear impact of the grid on the charge deposited on the pads.
Most important is therefore the alignment of the vertical bars and the pad plane.
By e.g. staggering the pads under the bar, it can be ensured that not a complete column of pads is shadowed by the grid.
The horizontal bars should be aligned to the middle of a pad row in order to minimize their impact.
For a large scale TPC, the design of the grid has to be adapted to the module layout to ensure that all pads are able to provide useful charge signals, which is desirable to preserve the good pattern recognition performance of the TPC.
In summary, with a proper design of the system the impact of the grid on the charge determination can be minimized.

\section{Grid Impact on Hit Reconstruction and Hit Efficiencies}
\begin{figure}
\unitlength1cm
\centering
\begin{picture}(15,7.5)
 \put(2,0.5){\includegraphics[height=6.5cm,clip=]{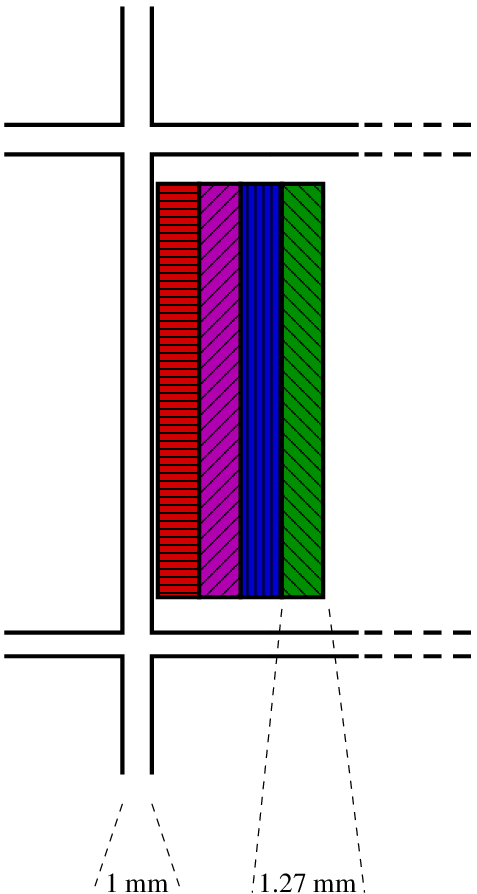}}
 \put(7.5,0){\includegraphics[width=7.5cm,clip=]{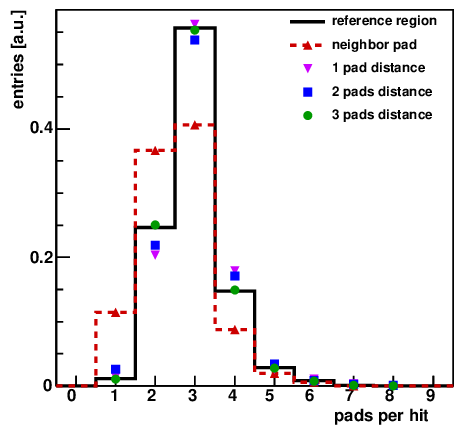}}
  \put(0,7.2){(a)}
 \put(7.5,7.2){(b)}
 \end{picture}
\caption{(a) Sketch of the neighboring vertical regions used for comparative studies of the influence of vertical grid bars.
Each region has a width of \unit[1.27]{mm}.
(b) Influence of the vertical grid bar on the number of pads per hit for the different regions, measured at a magnetic field of \unit[4]{T}. The color code used for the points is the same as used in the sketch on the left.
The distributions are normalized to the number of entries.}
\label{fig:vhit}
\end{figure}

In this section, the impact of the grid on the number of pads contributing to a hit and on the single hit efficiency is presented.
Hits are reconstructed by combining neighboring pads within a row, as described in section \ref{sec:setup}.
In figure \ref{fig:Qsumhhit}(b) and \ref{fig:vhit}(b), the average number of pads per hit is shown, studying the influence of horizontal (see figure \ref{fig:Qsumhhit}(b)) and vertical (see figure \ref{fig:vhit}(b)) grid bars.
In the solid histogram in both figures the distribution for hits where the impact of the grid bar is negligible is shown, in figure \ref{fig:Qsumhhit}(b) hits that are located more than \unit[5]{mm} from a grid bar and in figure \ref{fig:vhit}(b) more than one row.
In both cases, the average number of pads contributing to a hit is reduced close to the grid bars. As expected, the effect is more pronounced for vertical bars than for horizontal bars. The effect extends beyond the immediately affected pad.
%%%%%%%%%%%%%%%%%%%%%%%%%%%%%%%%%%%%%%%%%%%%%%%%%%%%%%%%%%%%%%
%%%%%%%%%%%%%%%%%%%%%%%%%%%%%%%%%%%%%%%%%%%%%%%%%%%%%%%%%%%%%%
For the horizontal bars, the number of pads per hit is reduced by \unit[5.0$\pm$1.0]{\%}, if the center of gravity of the hit is on top of the grid. For hits positioned two rows away no impact is seen. 
For the vertical bars, the number of pads per hit is reduced by \unit[13.9$\pm$1.0]{\%}. In both cases the error quoted is purely statistical. For hits with the reconstructed position 3 pads away, no effect is visible. 
Since the pad response function (PRF) has a width of \unit[0.67-0.78]{mm} for different drift distances, it can be stated that hits occurring more than two widths of the PRF ---or one pad pitch--- away from the grid structure are not influenced by the grid.

The changes on the number of pads contributing to a hit will have an impact on the hit reconstruction efficiency, and thus might ultimately have an impact on the efficiency with which tracks can be reconstructed. 

In the following, the efficiency to reconstruct a hit is defined as the number of reconstructed hits relative to the expected number of hits at this position. 
The study was performed on a sample of about 42,000 single cosmic muon tracks.
To determine the expected number of hits at a given position, tracks are searched for in the sensitive volume.
The row for which the hit efficiency is investigated is excluded from the track finding and fitting.
The expected hit position in the row under investigation is calculated from the parameters of the track.
A hit reconstructed in this row is tagged as belonging to the track if it is located within one pad width of the expected hit position.

With the used setup, tracks could have at most 10 rows contributing.
To ensure a sample of well defined tracks, all rows except the one under investigation are required to show a hit on the track. 
%%%%%%%%%%%%%%%%%%%%%%%%%%%%%%%%%%%%%%%%%%%%%%%%%%%%%%%%%%%%%%
\begin{figure}
\unitlength1cm
\centering
\begin{picture}(15,7.5)
 \put(0.1,0.15){\includegraphics[width=7.5cm,clip=]{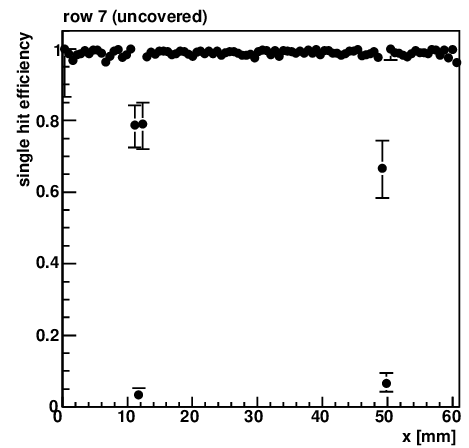}}
 \put(7.7,0){\includegraphics[width=7.5cm,clip=]{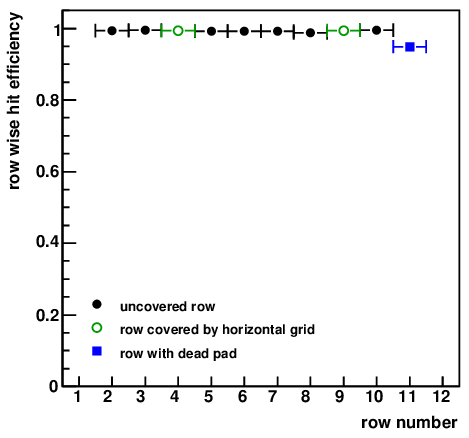}}
 \put(0.1,7.2){(a)}
 \put(7.7,7.2){(b)}
\end{picture}
\caption{(a) Single hit efficiencies for a row not covered by a horizontal grid bar.
The binning corresponds to half a pad pitch.
(b) Single hit efficiency as function of the row number.
An almost flat distribution can be observed.
Only the last row has a slightly lower value due to dead pads in this row.
All data measured at a magnetic field of \unit[4]{T}.}
\label{fig:hiteff}
\end{figure}
%%%%%%%%%%%%%%%%%%%%%%%%%%%%%%%%%%%%%%%%%%%%%%%%%%%%%%%%%%%%%%
The hit efficiency as a function of x for row seven (not influenced by a horizontal grid bar) is shown in figure \ref{fig:hiteff}(a).
The binning is chosen such that each bin corresponds to half a pad pitch, since the row-wise staggering corresponds exactly to this half pad pitch.
A drop in the efficiency is clearly visible for the x regions affected by the vertical grid bars. %For the affected pads, the efficiency drops around \unit[15$\pm$5]{\%}.
At directly covered pads the efficiency drops to almost zero; at neighboring pads, the efficiency drops about \unit[25$\pm$4]{\%}.
%, while the overall hit efficiency for this row can still be quoted with \unit[98]{\%} \cite{lea}.\\
The effects of the horizontal bars are studied with a sample of tracks where hits are at least 3 pad rows away from any vertical grid.
In figure \ref{fig:hiteff}(b) the hit efficiency is shown as a function of the row number, integrating over all x.
The impact of the horizontal grids in row four and nine is negligible, as expected.

\section{Impact of the Grid on the Track Reconstruction}
\label{sec:track}

Particle tracks are reconstructed from measured space points by fitting track parameters to these points.
In this section, the impact of the grid structures on the point resolution and possible biases in the reconstruction of the points due to the grid are studied.

\paragraph{Reconstruction bias}
The observed reduction of charge close to the grid can induce systematic shifts of the reconstructed hit position for hits close to a grid bar.
The bias is studied in a sample of reconstructed tracks.
The intercept between the track and the row position for the row investigated is calculated.
The distance between this intercept and the measured hit position on this row is studied.
For this study the hit investigated is included in the fit. 
%%%%%%%%%%%%%%%%%%%%%%%%%%%%%%%%%%%%%%%%%%%%%%%%%%%%%%%%%%%%%%
\begin{figure}
\unitlength1cm
\centering
\begin{picture}(15,7.5)
 \put(0.1,0){\includegraphics[width=7.5cm,clip=]{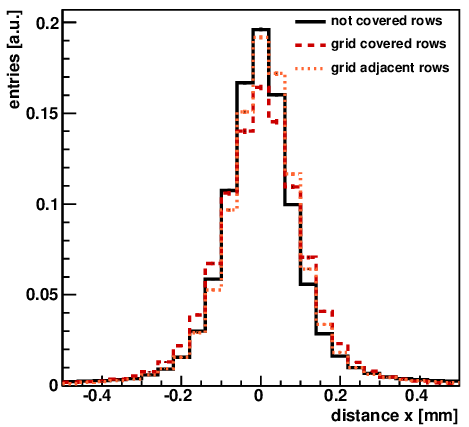}}
 \put(7.7,0){\includegraphics[width=7.5cm,clip=]{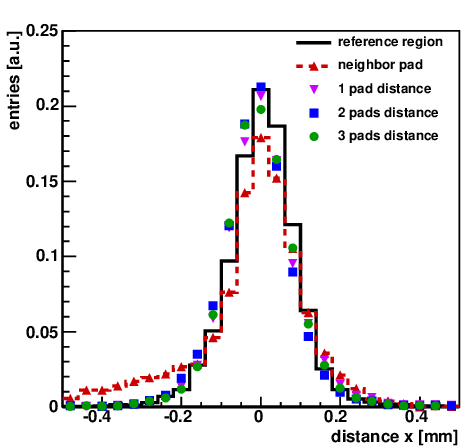}}
 \put(0.1,7.2){(a)}
 \put(7.7,7.2){(b)}
\end{picture}
\caption{Distribution of the mean values of hits reconstructed on tracks close to the horizontal and vertical bars. 
The data has been measured in a magnetic field of \unit[4]{T} and both distributions are normalized. 
(a) Influence of a horizontal grid bar on the hit distance in x direction.
(b) Impact of the vertical grid on the hit distance in x direction. Compared are regions with different distances to a vertical grid bar.
}
\label{fig:hitdist}
\end{figure}
%%%%%%%%%%%%%%%%%%%%%%%%%%%%%%%%%%%%%%%%%%%%%%%%%%%%%%%%%%%%%%
In figure \ref{fig:hitdist}(a), the impact of the horizontal grid is studied. 
The impact is very small, below \unit[10]{$\mu$m,} and within the statistical uncertainty. 
In figure \ref{fig:hitdist}(b) the impact of the vertical bar is shown. 
For the vertical bar, hits close to the bar are observed to be systematically shifted.
The distribution shows a significant tail.
The shift in this particular test system was found to be as large as \unit[40]{$\mu$m} for hits reconstructed at the position of the grid.
This is non negligible in size, but still small compared to the intrinsic resolution of the system.
The absolute size of the effect will depend on many factors, including drift distance, angle, the details of the readout, etc..

\paragraph{Single Point Resolution}
The single point resolution is of particular importance in tracking detectors.
It determines the momentum resolution of the TPC and gives a handle to judge the performance.\\
The single point resolution is calculated following the procedure outlined in~\cite{dean}.
To summarize briefly, for the hit under investigation the distance between this hit and the expected position of the track is calculated including and excluding the hit from the track fit.
The width of the distribution of the geometrical mean of these two numbers has been shown to be an unbiased estimator of the single hit resolution. 

%%%%%%%%%%%%%%%%%%%%%%%%%%%%%%%%%%%%%%%%%%%%%%%%%%%%%%%%%%%%%%
\begin{table}[h]
\centering
\caption{Track cuts for single point resolution determination.}
\frame{
\begin{tabular}{l r c l}
\multicolumn{1}{l}{variable} & \multicolumn{3}{c}{requirement}\\
\hline
number of tracks & \hspace{7mm}$\mr{n_{tracks}}$ & $=$ & 1\\
number of hits & $\mr{n_{hits}}$ & $=$ & $\mr{n_{rows}}$\\
x region & \multicolumn{3}{c}{$\mr{\unit[2.54]{mm} < x_{hit} < \unit[59.06]{mm}}$}\\
curvature & $\mr{\mid \kappa \mid}\hspace{0.4mm}$ & $<$ & \unit[0.02]{mm$^{-1}$}\\
inclination in yz-plane & $\mr{\mid \theta \mid }\hspace{0.4mm}$ & $<$ & \unit[0.45]{rad}\\
inclination in xy-plane & $\mr{\mid \phi \mid }$ & $<$ & \unit[0.1]{rad}\\
\end{tabular}
}
\label{tab:rescuts}
\end{table}
%%%%%%%%%%%%%%%%%%%%%%%%%%%%%%%%%%%%%%%%%%%%%%%%%%%%%%%%%%%%%%
The tracks are selected according to the cuts summarized in table \ref{tab:rescuts}.
Two samples of tracks are prepared which are sensitive to the impact of the horizontal and vertical bar, respectively. 
For the study of the horizontal bars, tracks with hits closer than three pad pitches to a vertical bar are excluded. 
The vertical bar is investigated by comparing tracks from the sample described above with those tracks that have been removed from this sample. 

\begin{figure}
\unitlength1cm
\centering
\begin{picture}(15,5.5)
  \put(0,0){\includegraphics[width=7.5cm,clip=]{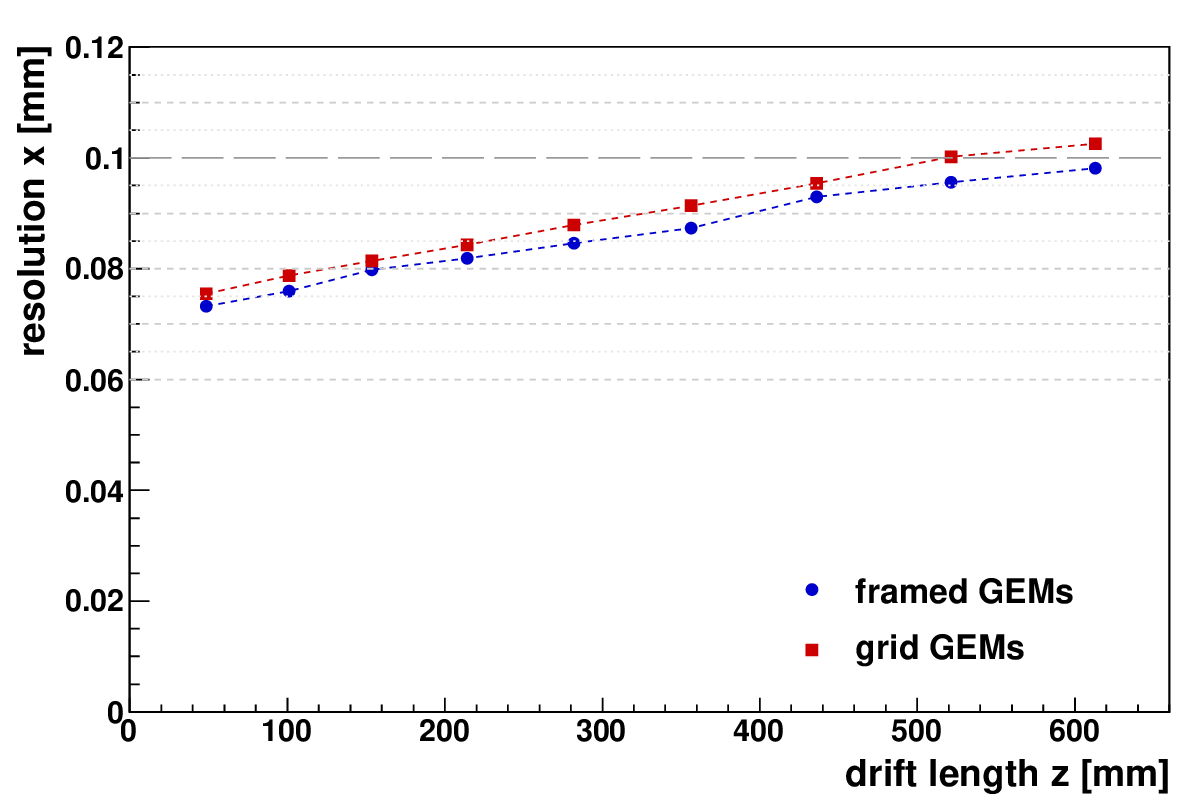}}
%  \put(7.7,0.0){\includegraphics[width=7.5cm,clip=]{2012_Resolution_Compare_Rowwise_Horizontal_largeLeg.eps}}
  \put(7.7,0){\includegraphics[width=7.5cm,clip=]{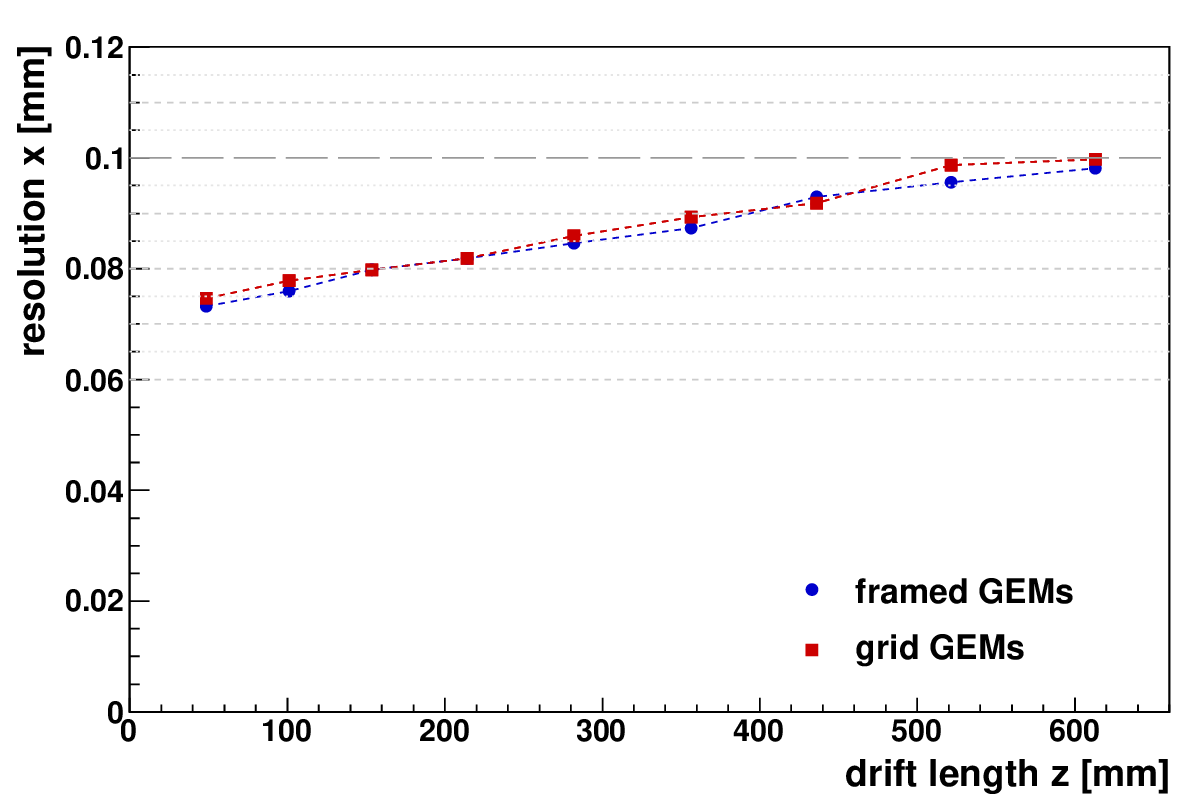}}
  \put(0,5.2){(a)}
 \put(7.7,5.2){(b)}
\end{picture}
\caption{
Single point resolution as a function of the drift distance, for data recorded with the grid GEM, and for data recorded with a standard framed GEM.
(a) No fiducial cuts are applied to the data.
(b) Hits which contain pads within 3 columns of a vertical bar are excluded from the sample. }

\label{fig:resallh}
\end{figure}

In figure \ref{fig:resallh}(a), the single point resolution for a sample of cosmic muons recorded with the TPC equipped with a grid GEM readout is shown.
For comparison results from a similar run but recorded with a GEM based readout with conventionally framed GEMS are also shown.
Both sets of data were taken at a magnetic field of \unit[4]{T} and with identical GEM settings.
In figure \ref{fig:resallh}(b) the same plots are shown, but this time only for tracks which do not contain hits within 3 columns of a vertical bar.
These two sets of plots clearly demonstrate that the horizontal bars do not have a significant impact on the resolution, while hits close to a vertical bar are found to have a worse single point resolution. 

%%%%%%%%%%%%%%%%%%%%%%%%%%%%%%%%%%%%%%%%%%%%%%%%%%%%%%%%%%%%%%
%%%%%%%%%%%%%%%%%%%%%%%%%%%%%%%%%%%%%%%%%%%%%%%%%%%%%%%%%%%%%%

\begin{figure}
\unitlength1cm
\centering
\begin{picture}(15,5.5)
%  \put(0,0){\includegraphics[width=7.5cm,clip=]{2012_Resolution_Compare_gridframe_largeLeg.eps}}
  \put(4,0.0){\includegraphics[width=7.5cm,clip=]{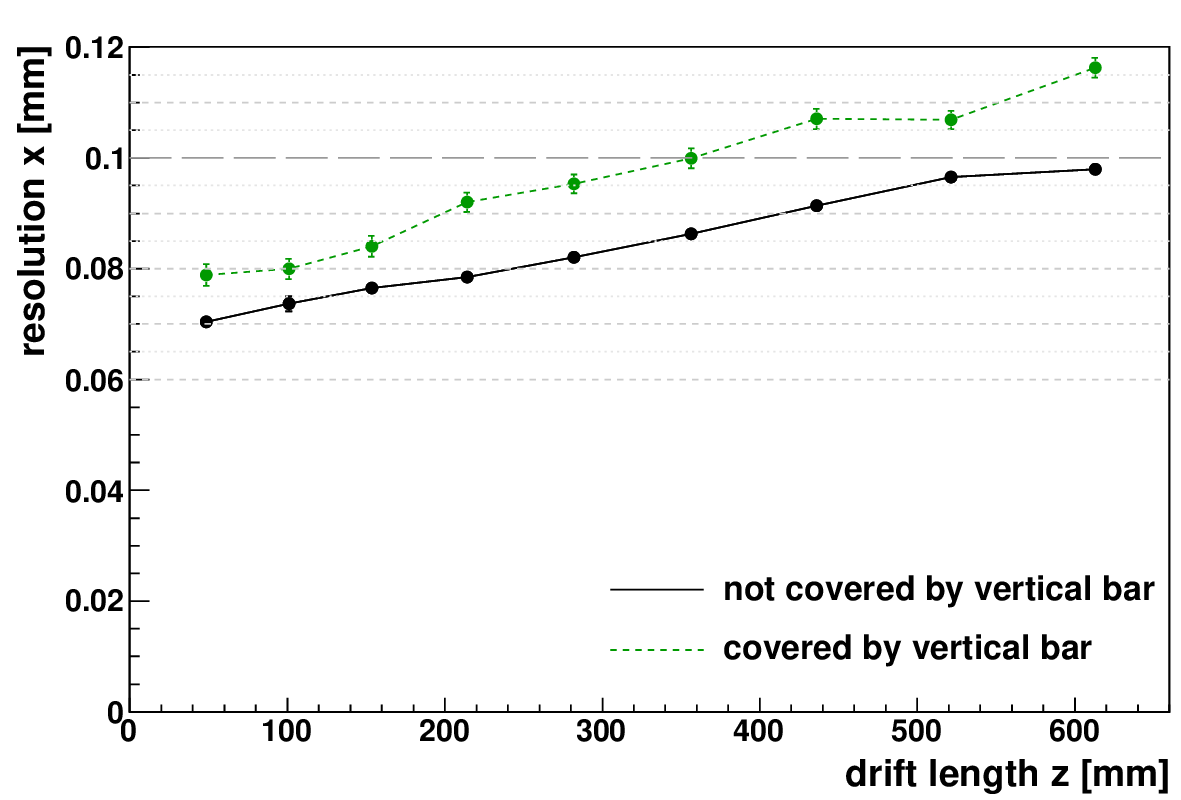}}
  %\put(0,5.2){(a)}
% \put(7.7,5.2){(b)}
\end{picture}
\caption{
Single point resolution for tracks with hits centered on pads close to or shadowed by a vertical bar, and for tracks with only hits far away from a grid bar.
Data recorded at a \unit[4]{T} magnetic field. 
%
%regions covered by a vertical grid bar (dashed green) and x ranges, where no vertical bar is shadowing the pad plane (solid black), measured at a magnetic field of \unit[4]{T}.
}
\label{fig:reshv}
\end{figure}

The effect is studied in more detail by comparing a sample of tracks close to and crossing a vertical bar with a sample of tracks in a reference region.
The single point resolution is calculated individually for tracks with hits close to (less than three pad pitches) and therefore affected by a vertical grid, and for tracks in the reference region far from both vertical grid bars.
The results are shown in figure~\ref{fig:reshv}.
The resolution deteriorates by about 20\% compared to all other hits. 

%%%%%%%%%%%%%%%%%%%%%%%%%%%%%%%%%%%%%%%%%%%%%%%%%%%%%%%%%%%%%%
%%%%%%%%%%%%%%%%%%%%%%%%%%%%%%%%%%%%%%%%%%%%%%%%%%%%%%%%%%%%%%
%\newpage
\section{Conclusion}
A novel scheme to mount GEM foils inside a TPC has been developed.
This self-supporting structure is made of a ceramic grid glued on the GEM foils.
The new support structure has been demonstrated to result in a stable, flat system which can be operated stably over extended periods of time. 

In order to quantify the impact of the grid GEMs on the track reconstruction, cosmic muon tracks have been recorded in a magnetic field of \unit[4]{T}.
The results show that the main impact of the grid is a reduction in the reconstructed charge on all pads which are shadowed by the grid. The effect depends on the fraction of a pad covered by the grid. It has been shown that the effects can be controlled at a level below the intrinsic hit resolution.

The advantages of the developed grid support structure are the minimal amount of material, the achievable flatness without the need of stretching the foils, the almost edgeless module borders and the resulting possibility to cover large areas without significant gaps.

\acknowledgments
This work was supported by the Commission of the European Communities under the 6$^\mathrm{th}$ Framework Programme 
'Structuring the European Research Area', contract number RII3-026126.
We would like to express our thanks to the DESY cryogenic operations team for the efficient operation of the 4T magnet facility.

\end{document}